%% file: sample-sigconf.tex
\documentclass[sigconf]{acmart}

\usepackage{booktabs} % For formal tables
\usepackage[hang]{subfigure}
\usepackage{balance}
\setcopyright{none}
\renewcommand\footnotetextcopyrightpermission[1]{} % removes footnote with conference information in first column
\begin{document}
\title{Plane-Casting: 3D Cursor Control with a SmartPhone}

\author{Nicholas Katzakis}
\orcid{0000-0002-1120-5648}
\affiliation{%
  \institution{Osaka University}
  \streetaddress{1-32 Machikaneyama}
  \city{Toyonaka}
  \state{Osaka}
  \postcode{560-0043}
}
\email{katzakis@lab.ime.cmc.osaka-u.ac.jp}

\author{Kiyoshi Kiyokawa}
   \affiliation{%
  \institution{Osaka University}
  \streetaddress{1-32 Machikaneyama}
  \city{Toyonaka}
  \state{Osaka}
  \postcode{560-0043}
}
    \email{kiyo@acm.org}
  
\author{Masahiro Hori}
\affiliation{
    \institution{Kansai University}
    \streetaddress{2-1-1, Ryouzenji-cho}
    \city{Takatsuki}
    \state{Osaka}
    \postcode{569-1095}
    }
\email{horim@res.kutc.kansai-u.ac.jp}

\author{Haruo Takemura}
   \affiliation{%
  \institution{Osaka University}
  \streetaddress{1-32 Machikaneyama}
  \city{Toyonaka}
  \state{Osaka}
  \postcode{560-0043}
}
\email{takemura@cmc.osaka-u.ac.jp}

\begin{abstract}
We present Plane-Casting, a novel technique for 3D object manipulation from a distance that is especially suitable for smartphones. We describe two variations of Plane-Casting, Pivot and Free Plane-Casting, and present results from a pilot study. Results suggest that Pivot Plane-Casting is more suitable for quick, coarse movements whereas Free Plane-Casting is more suited to slower, precise motion. In a 3D movement task, Pivot Plane-Casting performed better quantitatively, but subjects preferred Free Plane-Casting overall.
\end{abstract}

\maketitle

\input{samplebody-conf}

\bibliographystyle{ACM-Reference-Format}
\bibliography{arxiv}

\end{document}

%% file: samplebody-conf.tex
% =============================================================================
\section{Introduction}
% =============================================================================

3D interaction is a challenging problem and has been for over half a century, ever since the creation of the first 3D computer graphics. As hardware technology advanced, and display sizes grew, it became possible to view graphics from a distance, so the need to interact also ensued. However, currently available controllers for remote 3D control, among other weaknesses, lack in intuitiveness. Therefore, in this work we propose the use of smartphones as 3D controllers. State of the art smartphones feature an array of orientation sensors which make it possible to calculate the device's orientation in 3D. By further employing the touch-screen we demonstrate that with our proposed technique, Plane-Casting, it is possible to translate an object in 3D. In Plane-Casting, the rotation of the smartphone controls a virtual plane that constrains the movement of the 3D cursor. Aside from the potential for intuitive 3D control, the very wide availability of smartphones is an additional motivating factor for our work.

Examples of situations where there is a need to interact in 3D from a distance include the following: 
\begin{itemize}
\item Entertainment: As the number of displays in urban spaces increase, there are numerous opportunities for social entertainment that involve 3D (like 3D Games). 
\item Design: A team of designers is reviewing the latest 3D assets in their weekly meeting. Participants interact, review and discuss changes relating to the 3D geometry.
\item Education: A medical school professor is demonstrating the anatomy of the human heart by projecting 3D graphics. A smartphone controller allows the professor to leave the podium and approach the students while still being able to interact with the model, thus making the class more engaging. Students can also use their smarphones to actively participate.
\end{itemize}

In the remainder of this paper we present two variations of Plane-Casting, \emph{Pivot} Plane-Casting and \emph{Free} Plane-Casting. We discuss their strengths and limitations and present results from a pilot study.

\begin{figure*}
\centering  
\subfigure[hang][Pivot PC: The user gestures to translate the acquired object which moves on the plane.]{\label{fig:ppc1}\includegraphics[width=0.32\linewidth]{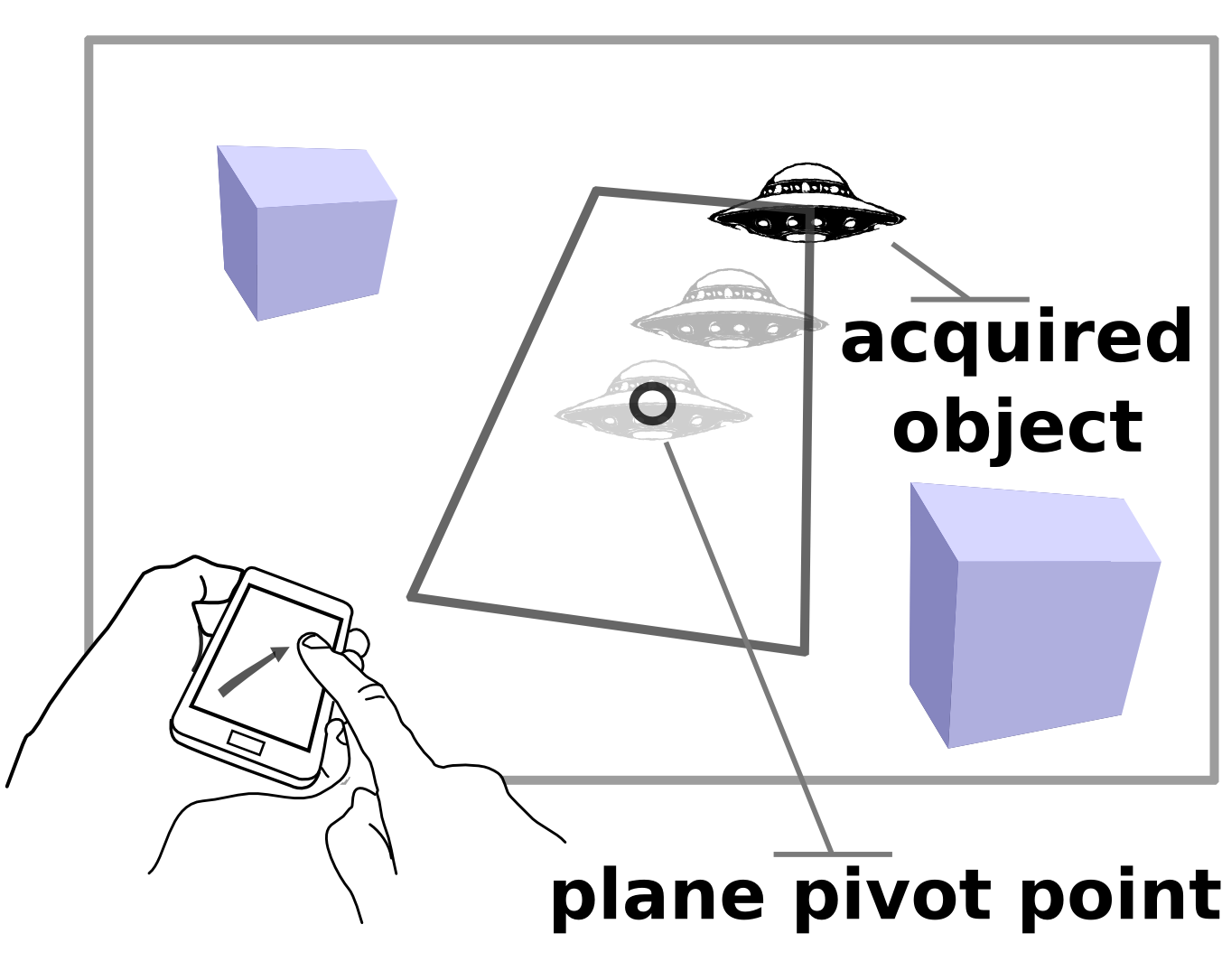}}
\subfigure[hang][Plane follows rotation of device. Rotates about the pivot with the object bound to it.]{\label{fig:ppc2}\includegraphics[width=0.32\linewidth]{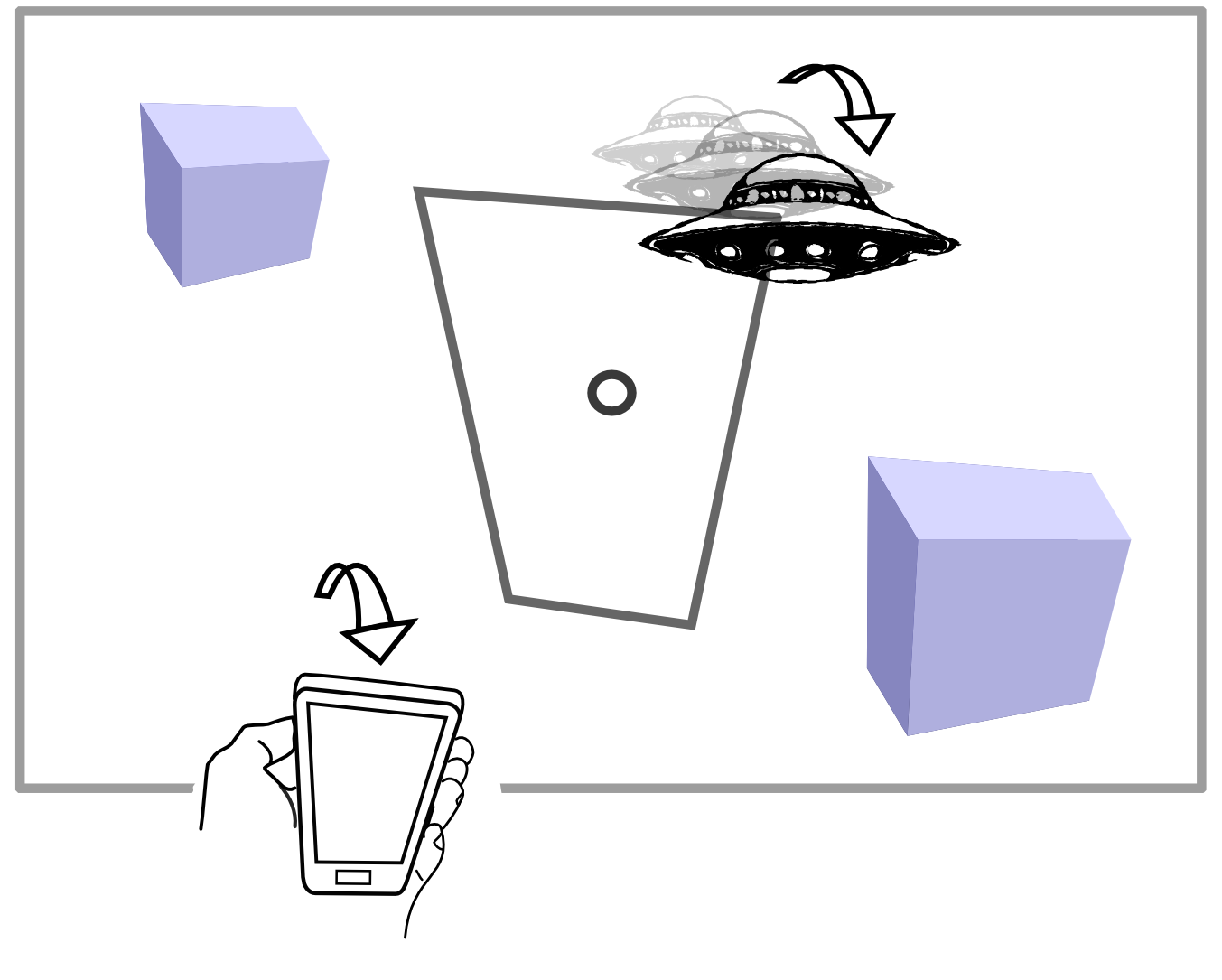}}
\subfigure[hang][The object's rotation does not change, only it's position.]{\label{fig:ppc3}\includegraphics[width=0.32\linewidth]{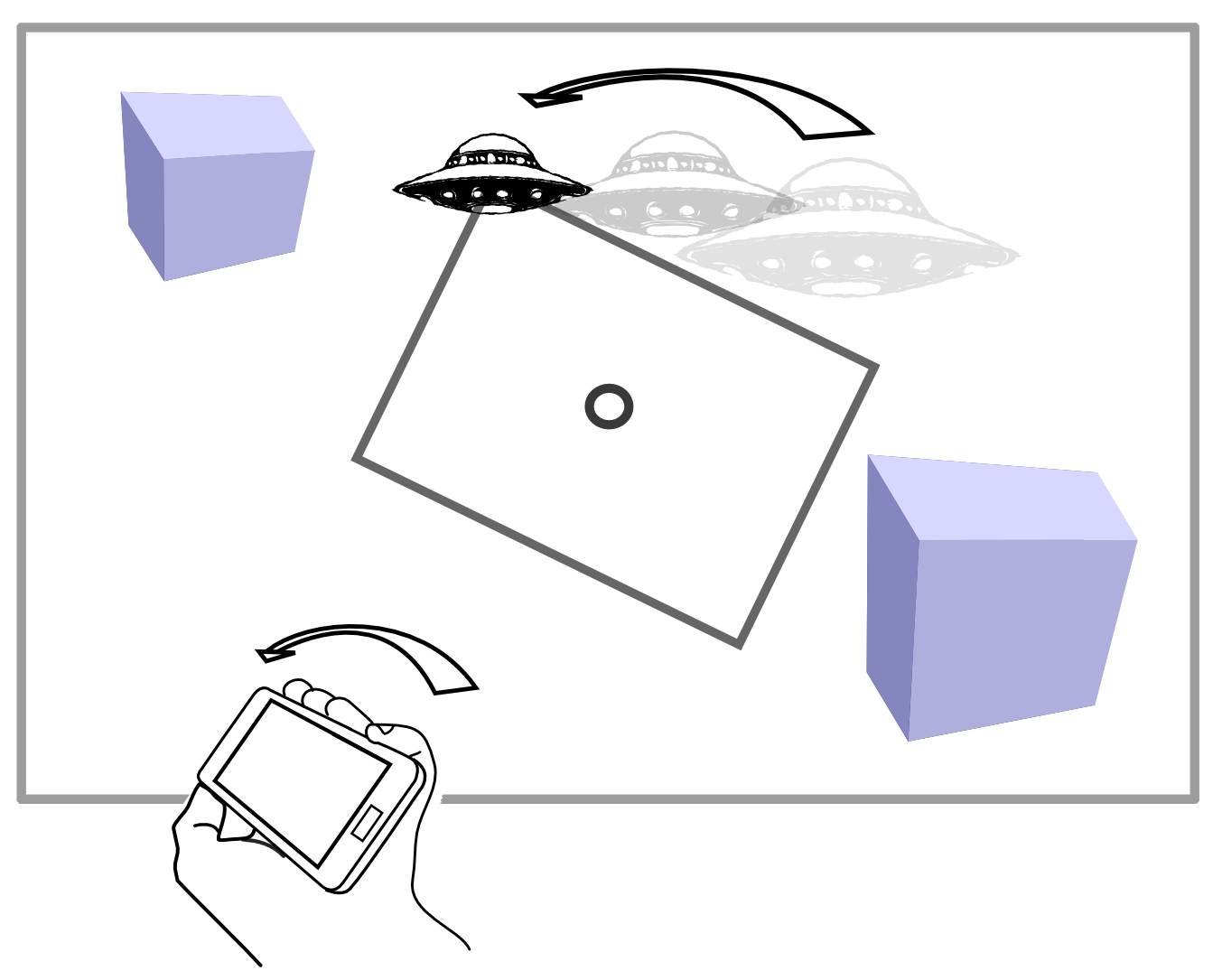}}

\subfigure[hang][Free PC: The acquired object moves on the plane but the plane stays attached to it.]{\label{fig:fpc1}\includegraphics[width=0.32\linewidth]{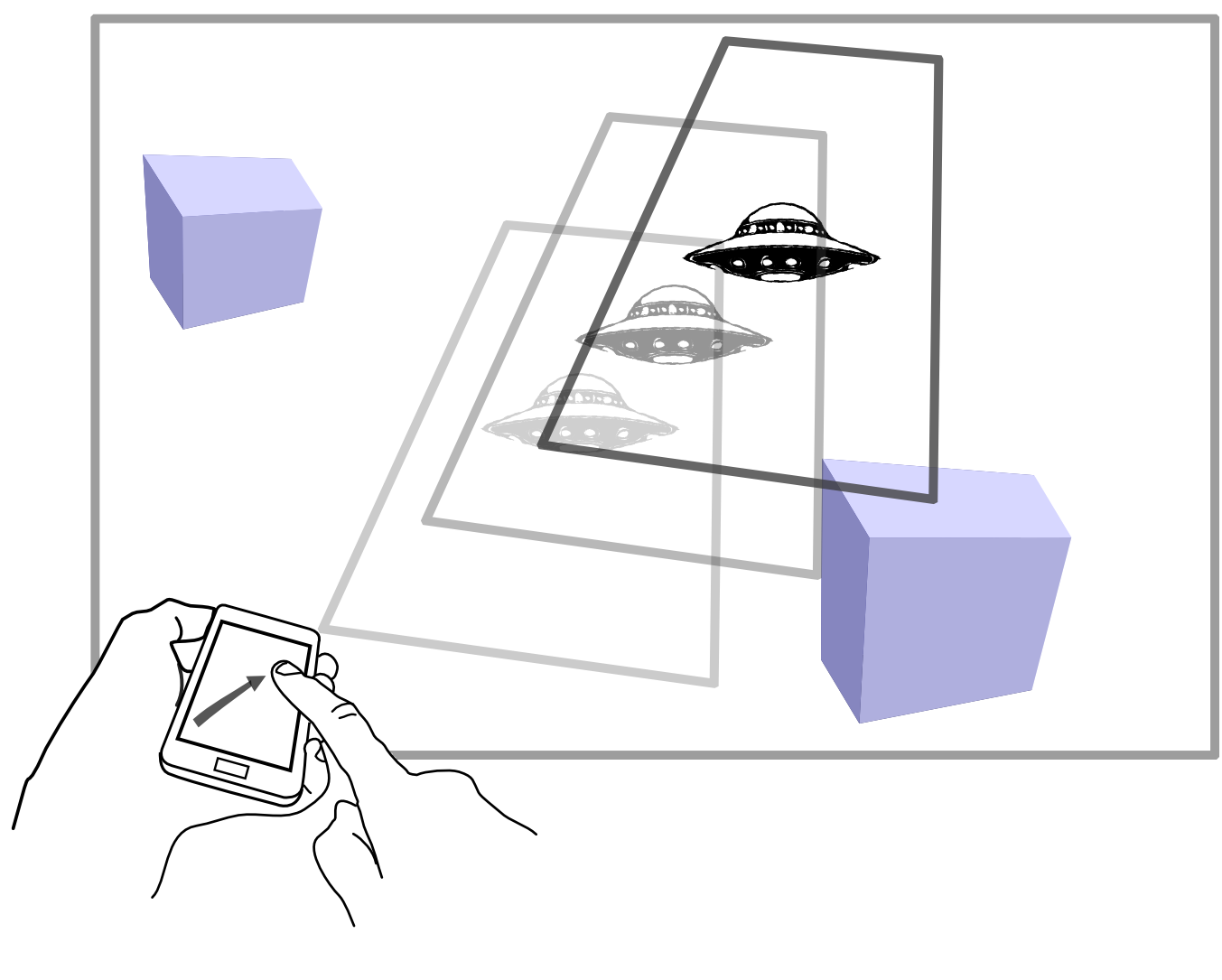}}
\subfigure[hang][The pivot point of the plane is always fixed at the center of the object's bounding box.]{\label{fig:fpc2}\includegraphics[width=0.32\linewidth]{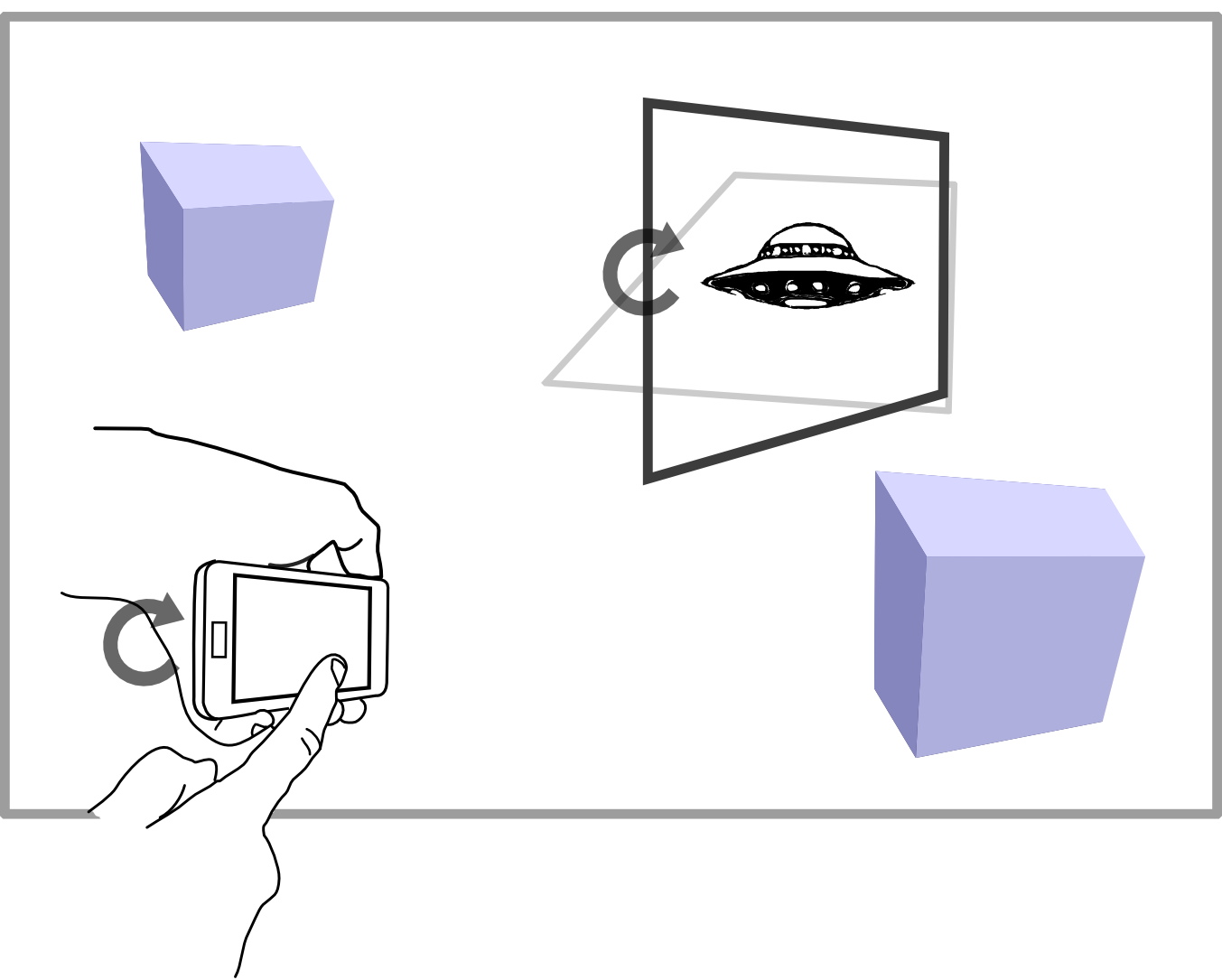}}
\subfigure[hang][Gesturing towards the motion direction regardless of the device's orientation.]{\label{fig:fpc3}\includegraphics[width=0.32\linewidth]{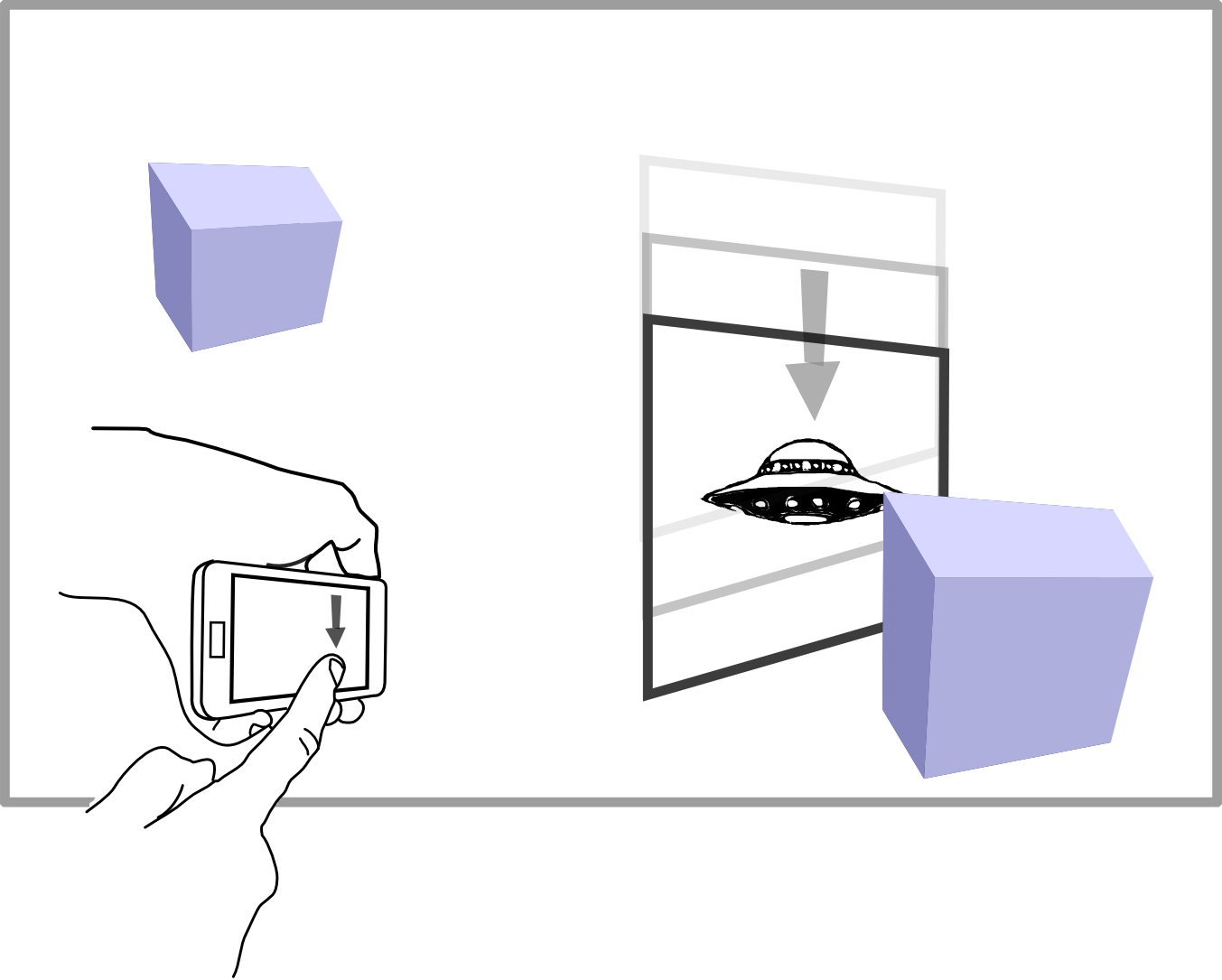}}

  \caption{The two variations of Plane-Casting}
  \label{fig:planecasting}

\end{figure*}

\section{Related Work}
Touch-input enables users to interact with a display by removing an indirection layer, and there are quite a few solutions for 3D interaction by using multi-touch\cite{multitouch}. However when the display size exceeds a certain threshold, touch input ceases to be an option as the user needs to cover a large area with physical movements and in some cases the display area is out of reach (as is the case with projectors and Tile-displays). In addition to the input problems of touch, physically approaching the display to interact limits the user's activity to a very small area of the display and in the case of collaborative work, the interacting user obscures the display for the rest of the group. As an alternative to touch, gesture approaches require a carefully controlled environment and slightly lack in efficiency for practical use. 

The Nintendo Wii-mote\texttrademark{} is a popular choice for remote control but depends on a 2-state directional-pad for additional degrees of freedom. Other controllers like the 3Dconnexion SpaceNavigator\texttrademark{} depend on desks and are tethered by cables, thus making them unsuitable for an active, engaging experience or for use in public or shared spaces. 

More directly related to our work, Bier's discussion on constraining motion in a scene composition scenario is one of the earliest references in the literature \cite{bier86}. Bier further emphasizes the power of constraint-based systems in subsequent works.

Hachet et al.\cite{HatchetElastic2008} propose a controller that attaches to the side of mobile phones and can provide 3-DOF control, a solution which could be used for remote 3D control. They evaluate the design in a navigation scenario and report positive reactions from the users. Their approach is however based on proprietary hardware external to the device, and limited to rate control.

More recently, Jimenez et al.\cite{mobilemuseum} used a hand-held device in a museum scenario for remote assembly of a puzzle-like task. Their work highlights some of the social aspects of using a hand-held interface in a collaborative task. Their evaluation suggests that a usable interface might better promote equal participation in a group task.

Finally, Song et al.\cite{planechi} used a hand-held device in a large-display scenario. The device controls the position of a slicing plane in 5-DOF that explores volume-rendering data and the authors present a novel technique to annotate them. Song's approach unfortunately requires physical proximity to the screen which makes it unsuitable for remote or collaborative work and also depends on proprietary hardware attached to the hand-held device thus limiting it's applicability. Their paper offers a thorough review of the literature on hand-held/remote interaction.

Although there are a few interaction techniques that use magnetic trackers like the go-go technique\cite{gogo} or WIM\cite{wim}, there is currently no established standard technique/device for remote graphics manipulation. Although smartphones have been used in the past, solutions have been inadequate.

\section{Plane-Casting}

\subsection*{Pivot Plane-Casting}

In \emph{Pivot Plane-Casting} (PivotPC) the shape of the touch-screen is drawn as a rectangle at the center of the scene (Figure \ref{fig:ppc1}). The user can rotate the device to control the orientation of the plane (position-controlled). The plane's pivot point is at the center of the rectangle and always remains fixed at the center of the virtual space. The 3D cursor's movement is constrained on the plane defined by the rectangle but not limited to it's bounds. The user can translate the cursor on the plane by gesturing on the touch-screen and rotate the plane to move the cursor to any point in 3D space (Figure~\ref{fig:ppc2}). In our implementation the tactile-sensor is a touch screen but any touch panel that can be tracked is suitable for Plane-Casting.

\begin{figure}
\begin{center}
\includegraphics[width=0.7\linewidth]{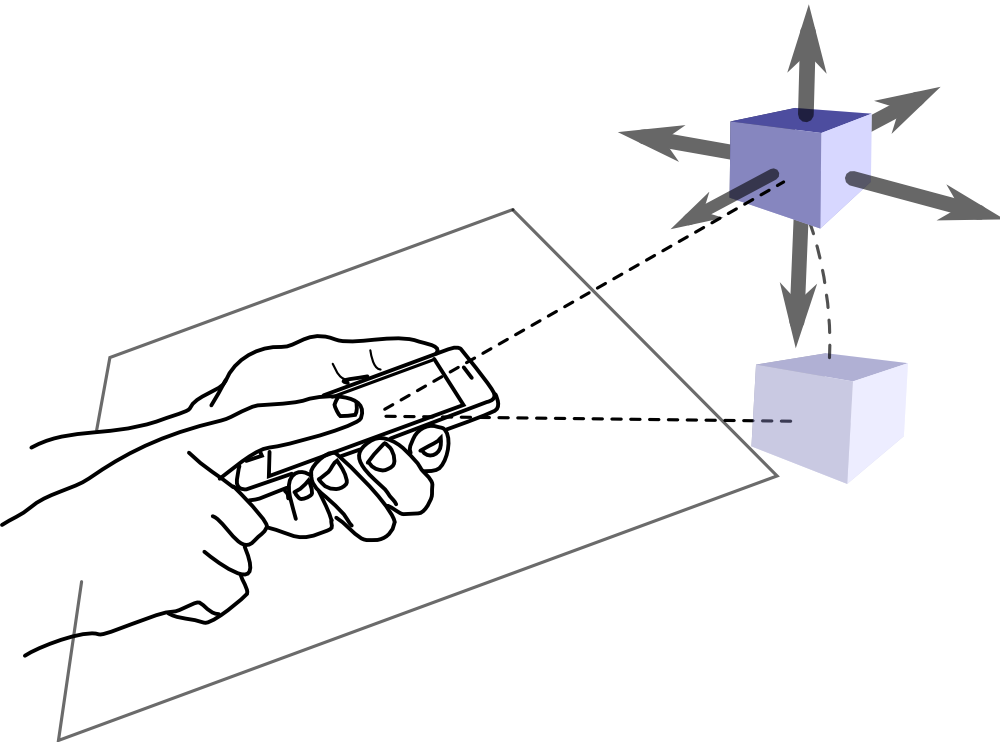}
\end{center}
\caption{PivotPC, moving vertically to the plane becomes easier as the object moves away from the pivot point of the plane.}
\label{fig:dof}
\end{figure}

%\marginpar{
%\begin{figure}
%  \begin{center}
%  \includegraphics[width=\marginparwidth]{sample.jpg}
%  \caption{A marginal figure.}
%  \label{fig:marginparsample}
%  \end{center}  
%\end{figure}
%}

\subsection*{Free Plane-Casting}

\emph{Free Plane-Casting} (FreePC) is similar to PivotPC but in this variation the plane's pivot point follows the cursor's motion in 3D space. FreePC shifts the center/pivot point around with evey slide movement. The rectangle that defines the plane is thus always attached to the cursor that is being manipulated and they move as one, with the orientation of the rectangle constantly re-defining the plane (Figure \ref{fig:fpc1}).

In PivotPC, placing the cursor away from the pivot point of the plane makes it easier to rotate the object vertically, on the normal to the current plane thus making it easier to quickly change direction as would be the case in a game (Figure \ref{fig:dof}), yet by sacrificing accuracy. FreePC's nature makes it so only 2-DOF are instantly available at any time and moving in a direction on the normal to the current plane requires a small supination/pronation move (Figure \ref{fig:fpc2}).

In our implementation of FreePC and PivotPC selection of the object to be manipulated is done by a spherical cursor that intersects the desired object in a widely used "virtual hand" metaphor. Depending on the application there are many strategies for object selection but that remains beyond the scope of this work.

\section{Evaluation}
Our pilot study evaluated the two techniques against each other in a 3D positioning task. A remote manipulation scenario like the ones mentioned in the introduction would require the ability to move an object in 3D, but which technique would be best for this? We wanted to test the following two hypotheses:
\begin{itemize}
\item H1 : PivotPC will perform faster than FreePC since it only requires an initial alignment of the plane to the target.
\item H2 : FreePC will be more accurate as it's essentially bringing the pivot point closer to the target and does not require a "steady hand" like PivotPC.
\end{itemize}

12 right-handed male participants, students and faculty volunteered for the experiment (age mean 25). Participants had no prior experience in using the techniques.

\subsection*{Set-up}
\begin{figure}
\includegraphics[width=\linewidth]{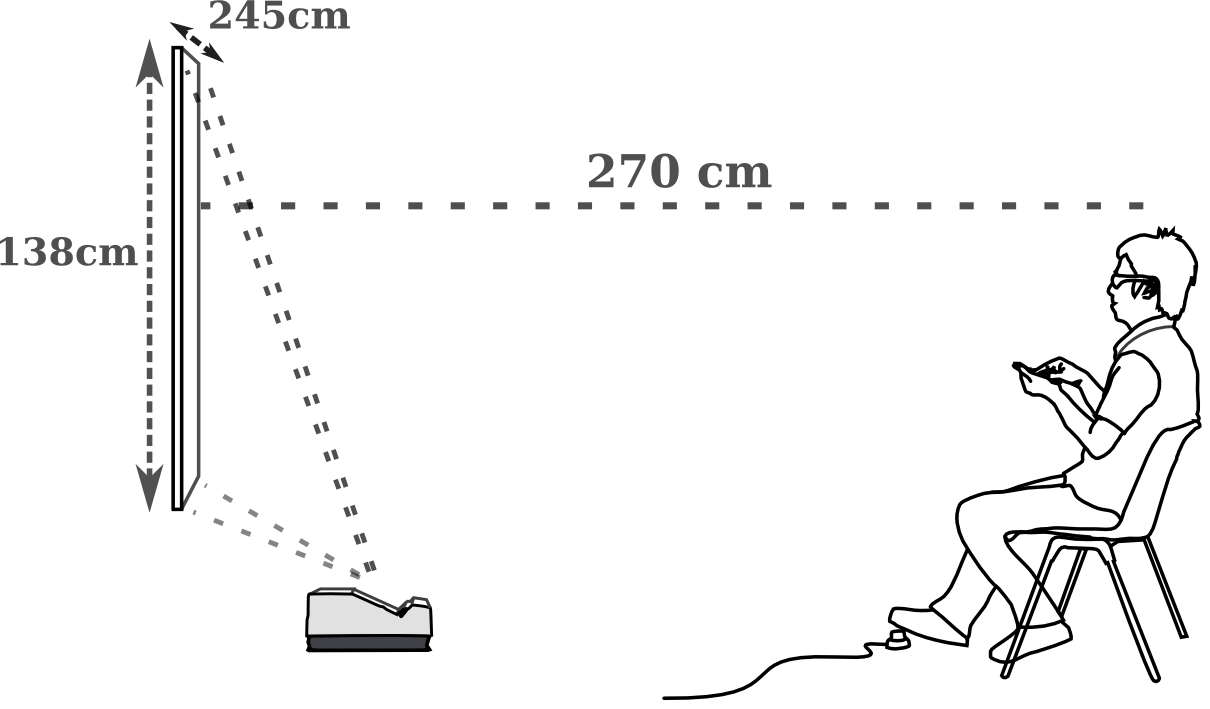}
\caption{Illustration of the experimental setup}
\label{fig:experiment}
\end{figure}

Subjects sat 270 cm from the projection screen of an ultra-short-focus projector (Sanyo PDG-DWL2500J). They were instructed to hold the smartphone (Samsung Galaxy SII) in their non-dominant hand while gesturing on the touch-screen with their dominant hand (Figure~\ref{fig:experiment}). A foot switch was available for advancing to the next trial. The projection screen had a width and height of 245x138cm respectively with a 1280x800 display resolution in stereoscopic 3D (Nvidia 3D Vision). Data transmission of the device sensor information was over an IEEE 802.11g WiFi link and was filtered with a 30 sample moving average filter for stabilization.

\begin{figure}
\includegraphics[width=\linewidth]{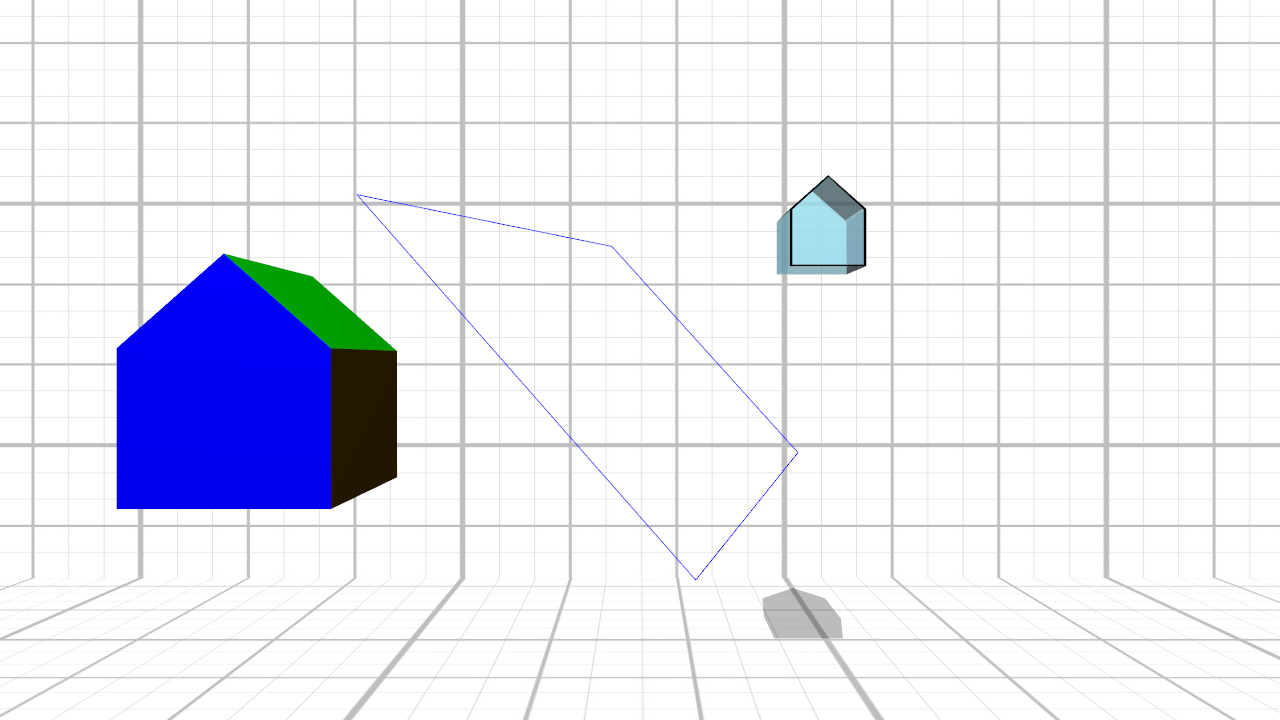}
\label{fig:screenshot}
\caption{Screenshot of the evaluation task (in monoscopic 3D) using PivotPC. Users had to dock the cursor (multi-colored-house) to the translucent target.}
\end{figure}

\subsection{Task}
In the evaluation task, the house-shaped cursor (and rectangle) appeared between the viewpoint and the far wall of the 3D space (Figure 4). When the experiment commenced, a translucent copy of the cursor appeared randomly at one of 12 pre-defined positions around the cursor (Figure \ref{targetlayout}) and subjects had to match the position of the cursor with that of the target under two conditions: 1) as quickly and 2) as  accurately as possible. The targets appeared in positions distributed evenly on the surface of a sphere centered at the cursor's starting position with a radius of either 52 or 96 cm respectively (Figure \ref{targetlayout}). Each position was tested twice with each technique, one in the \emph{Speed} and one in the \emph{Accuracy} condition (balanced order).

When the cursor's bounding box intersected the target's bounding box, the target's bounding box would become visible signaling a match. Subjects could not end the trial if they did not have a match. When subjects felt they had achieved a good match, they pressed the foot switch and the trial ended with both the cursor and the target disappearing. Only the rectangle remained. In FreePC, the pivot point of the plane/rectangle would return to the center of the scene. The next trial would only begin when subjects returned the device to it's original orientation parallel to the ground at which point the cursor would re-appear at the center of the plane, and the target at the next position to be tested. 

All subjects received a brief explanation of the techniques and performed the task once with each technique as practice (order of techniques was also balanced). Subjects performed 12 positions $\times$ 2 radii $\times$ 2 techniques $\times$ 2 (speed vs accuracy) = 96 trials (1152 total). The experiment lasted around 40 minutes with no break between conditions.

\section{Results - Discussion}
We recorded the movement time (MT)\footnote{Movement time started counting either when users rotated the device past a 2\% threshold - thus signaling their attempt to align the plane to the target - or when they first swiped on the touch-screen, whichever came first}, the accuracy as a measure of the euclidean distance (d) between the center of the cursor's bounding box and the target's bounding box at the time the subject pressed the foot-switch. We also recorded the distance the user's finger traveled on the touch screen (T)\footnote{The T measurement is the result of the total distance in pixels the finger traveled on the touch-screen while pressed down during the trial} during every trial.

\subsection{Technique Effect}
Results of a repeated-measures ANOVA showed that \emph{Technique} had a strong effect on movement time (MT). The average MT for FreePC was 9.7s vs 8.2s for PivotPC (F$_{(1,11)}$=17.1 p$<$.001). This confirms our first hypothesis (H1) of PivotPC being the quicker technique of the two, though based on the fact that FreePC requires repeated supination/pronation we expected the performance difference to be greater. 

Technique also had a significant effect on the amount participants gestured on the touch-screen. An average amount of ~190 pixels traveled was recorded when using PivotPC whereas when using FreePC users gestured an average of 225 pixels (F$_{(1,11)}$=23.6 p$<$.001). These results suggest that PivotPC is potentially suitable to implement on devices with smaller touch-surfaces than FreePC.

Contrary to our second hypothesis (H2), neither technique was found to be more accurate (d=6.5 for FixedPC vs 6.8 for FreePC F$_{(1,11)}$=0.3 p$=$0.58). 

\subsection{Target Distance Effect}
Whether the target was placed in the near radius 52cm or the far radius 96cm had no significant effect on the time cursors took to reach the target (F$_{(1,11)}$=3.2 p$<$0.09) with 8.9sec to reach the far ones vs 8.7 to reach the near ones. 

There was also no effect on the amount users gestured on the touch screen between radii (F$_{(1,11)}$=1.5 p$=$0.2) with users gesturing 203px for the near targets vs 210px for the far targets. 

Finally the distance of the target had no effect on accuracy as users were equally accurate on near and far targets (F$_{(1,11)}$=0.4 p$=$0.5).

\subsection*{Accuracy/Speed Tradeoff}
As would be expected, there was a strong effect on movement time and accuracy by the speed/accuracy condition. On the \emph{Speed} condition, users were asked to be as quick as possible with reasonable accuracy and on the \emph{Accurate} condition they were asked to be as accurate as possible again with reasonable speed. As such when MT is concerned users had an average of 5.2s in the \emph{Speed} condition vs 12.7s in the \emph{Accuracy} condition (F$_{(1,11)}$=115.8 p$<$0.001). Similarly accuracy (d) was 5.6 in the \emph{Accuracy} condition vs 8.7 in the \emph{Speed} condition (F$_{(1,11)}$=27 p$<$0.001).

\begin{figure}
\centering 
\includegraphics[width=\linewidth]{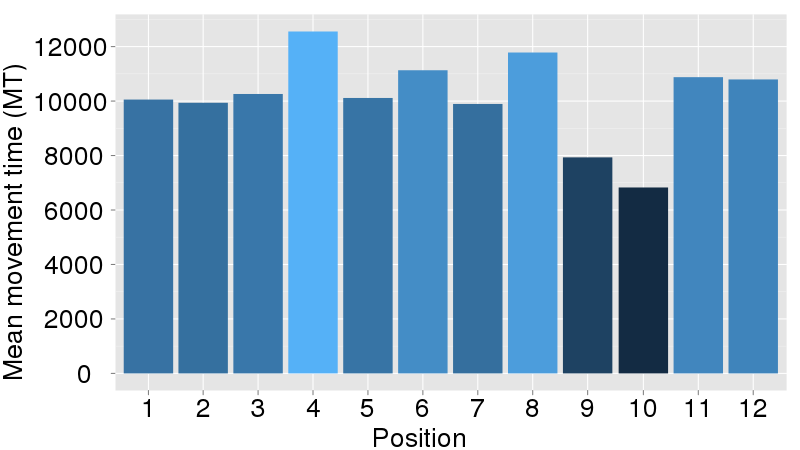}
\caption{FreePC: Mean movement time for every target position.}
\label{fig:fpc-pos-mt}
\end{figure}

\subsection{Target Position Effect}
\begin{figure}
\includegraphics[width=\linewidth]{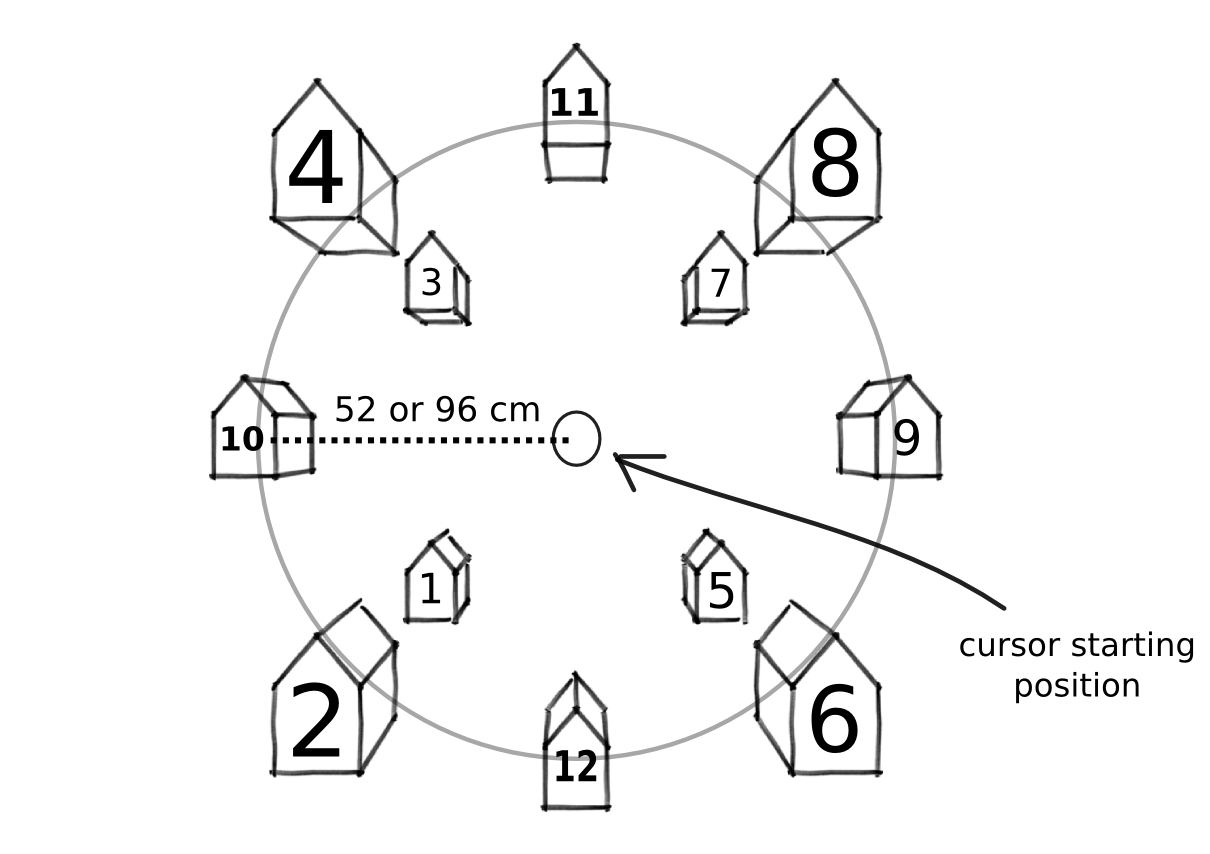}
\caption{Layout of the targets. 12 positions evenly distributed on a sphere around the cursor starting position with 4 of them axis-aligned. Targets 1-8 are pivoted 45$^\circ$ about the Y and Z axis. The two missing front and back axis-aligned positions were not tested since they occluded or were occluded by the cursor and slightly confused participants.}
\label{targetlayout}
\end{figure}

\begin{figure}
\centering
\includegraphics[width=\linewidth]{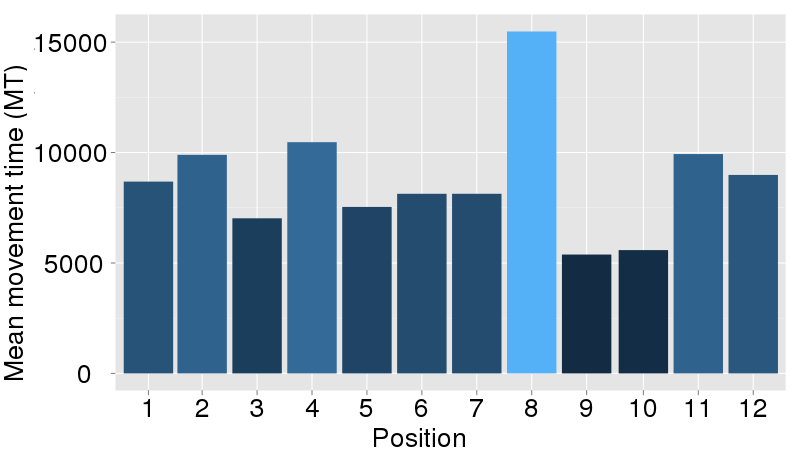}
\caption{PivotPC: Mean movement time for every target position. Position 8 was significantly slower to reach.}
\label{fig:ppc-pos-mt}
\end{figure}

Target position had a strong effect on MT. Particularly positions 9 and 10 which are horizontally aligned to the starting position were the fastest in both techniques, as expected (Figure~\ref{fig:fpc-pos-mt}-\ref{fig:ppc-pos-mt}).

\balance
In FreePC (Figure~\ref{fig:fpc-pos-mt}) there was a relatively even distribution of movement times between the various positions. In PivotPC (Figure~\ref{fig:ppc-pos-mt}), however, users struggled with position 8 (mean time ~15.5sec, (F$_{(11,121)}$=3.8 p$<$0.005).

Observations of the subjects indicate that the reason for the problem in position 8 was the limited ulnar and radial flexion of the human arm which makes it slightly strenuous to reach that position. One of the strengths of PivotPC is that there are many combinations of touch and rotation to reach the same position in 3D space. That is also part of it's weakness as it requires training to find which is the best combination of tilt/swiping to achieve the desired motion, something which, however, remains to be validated.

Finally, position of the target had no significant effect on accuracy (F$_{(11,121)}$=1.5 p$=$0.12) as users were equally accurate in all positions.

\subsection{Qualitative Results}
Subjects answered a post-experimental questionnaire asking them to rate PivotPC and FreePC in terms of their intuitiveness and physical demands. Summarizing the responses from the questionnaires shows that subjects clearly favor Free Plane-Casting over Pivot Plane-Casting both in intuitiveness and in physical demands, even if overall quantitative performance was better in PivotPC for this specific task. Subjects also commented that in FreePC, since the cursor can only move when gesturing, there is less "pressure" to keep the device aligned with the target (as is required in PivotPC - Figure \ref{fig:dof}) and that is cognitively (and physically) less demanding.

\section{Conclusion - Future Work}
We have introduced 2 variations of a novel technique for 3D cursor manipulation using a smartphone. Our pilot study verifies their usability and highlights some of the issues associated with each one. For the docking task PivotPC seems to be the quantitative overall winner, but subjects preferred FreePC. Further evaluation is required to ascertain their applicability to real-life scenarios. This work establishes a broad base upon which more specific plane-casting-based 3D applications can be built.

As future work we also plan to implement simultaneous rotation using multi-touch gestures as well as to develop techniques for annotation using touch.

%% file: sample-sigconf.bbl

\begin{thebibliography}{7}

%%% the Web address.

\ifx \showCODEN    \undefined \def \showCODEN     #1{\unskip}     \fi
\ifx \showDOI      \undefined \def \showDOI       #1{#1}\fi
\ifx \showISBNx    \undefined \def \showISBNx     #1{\unskip}     \fi
\ifx \showISBNxiii \undefined \def \showISBNxiii  #1{\unskip}     \fi
\ifx \showISSN     \undefined \def \showISSN      #1{\unskip}     \fi
\ifx \showLCCN     \undefined \def \showLCCN      #1{\unskip}     \fi
\ifx \shownote     \undefined \def \shownote      #1{#1}          \fi
\ifx \showarticletitle \undefined \def \showarticletitle #1{#1}   \fi
\ifx \showURL      \undefined \def \showURL       {\relax}        \fi
% invisible to TeX
\providecommand\bibfield[2]{#2}
\providecommand\bibinfo[2]{#2}
\providecommand\natexlab[1]{#1}
\providecommand\showeprint[2][]{arXiv:#2}

\bibitem[\protect\citeauthoryear{Bier}{Bier}{1986}]%
        {bier86}
\bibfield{author}{\bibinfo{person}{Eric~A. Bier}.}
  \bibinfo{year}{1986}\natexlab{}.
\newblock \showarticletitle{{Skitters and Jacks: Interactive 3D Positioning
  Tools}}. In \bibinfo{booktitle}{\emph{Proc. workshop on Interactive 3D
  graphics}}. \bibinfo{publisher}{ACM}, \bibinfo{pages}{183--196}.
\newblock
\showISBNx{0-89791-228-4}
\urldef\tempurl%
\url{https://doi.org/10.1145/319120.319135}
\showDOI{\tempurl}


\bibitem[\protect\citeauthoryear{Hachet, Pouderoux, and Guitton}{Hachet
  et~al\mbox{.}}{2008}]%
        {HatchetElastic2008}
\bibfield{author}{\bibinfo{person}{Martin Hachet}, \bibinfo{person}{Joachim
  Pouderoux}, {and} \bibinfo{person}{Pascal Guitton}.}
  \bibinfo{year}{2008}\natexlab{}.
\newblock \showarticletitle{{3D Elastic Control for Mobile Devices}}.
\newblock \bibinfo{journal}{\emph{IEEE Computer Graphics and Applications}}
  \bibinfo{volume}{28}, \bibinfo{number}{4} (\bibinfo{year}{2008}),
  \bibinfo{pages}{58--62}.
\newblock
\showISSN{0272-1716}
\urldef\tempurl%
\url{https://doi.org/10.1109/MCG.2008.64}
\showDOI{\tempurl}


\bibitem[\protect\citeauthoryear{Hancock, Carpendale, and Cockburn}{Hancock
  et~al\mbox{.}}{2007}]%
        {multitouch}
\bibfield{author}{\bibinfo{person}{Mark Hancock}, \bibinfo{person}{Sheelagh
  Carpendale}, {and} \bibinfo{person}{Andy Cockburn}.}
  \bibinfo{year}{2007}\natexlab{}.
\newblock \showarticletitle{Shallow-depth 3d interaction: design and evaluation
  of one-, two- and three-touch techniques}. In
  \bibinfo{booktitle}{\emph{CHI}}. \bibinfo{publisher}{ACM},
  \bibinfo{pages}{1147--1156}.
\newblock
\showISBNx{978-1-59593-593-9}


\bibitem[\protect\citeauthoryear{Jimenez and Lyons}{Jimenez and Lyons}{2011}]%
        {mobilemuseum}
\bibfield{author}{\bibinfo{person}{Priscilla Jimenez} {and}
  \bibinfo{person}{Leilah Lyons}.} \bibinfo{year}{2011}\natexlab{}.
\newblock \showarticletitle{{An exploratory study of input modalities for
  mobile devices used with museum exhibits}}. In
  \bibinfo{booktitle}{\emph{Proc. of conference on Human Factors in Computing
  Systems}}. \bibinfo{publisher}{ACM}, \bibinfo{pages}{895--904}.
\newblock
\showISBNx{978-1-4503-0228-9}
\urldef\tempurl%
\url{https://doi.org/10.1145/1978942.1979075}
\showDOI{\tempurl}


\bibitem[\protect\citeauthoryear{Poupyrev, Billinghurst, Weghorst, and
  Ichikawa}{Poupyrev et~al\mbox{.}}{1996}]%
        {gogo}
\bibfield{author}{\bibinfo{person}{Ivan Poupyrev}, \bibinfo{person}{Mark
  Billinghurst}, \bibinfo{person}{Suzanne Weghorst}, {and}
  \bibinfo{person}{Tadao Ichikawa}.} \bibinfo{year}{1996}\natexlab{}.
\newblock \showarticletitle{The go-go interaction technique: non-linear mapping
  for direct manipulation in VR}. In \bibinfo{booktitle}{\emph{Proc. of the 9th
  symposium on User interface Software and Technology}}
  \emph{(\bibinfo{series}{UIST '96})}. \bibinfo{publisher}{ACM},
  \bibinfo{pages}{79--80}.
\newblock
\showISBNx{0-89791-798-7}
\urldef\tempurl%
\url{https://doi.org/10.1145/237091.237102}
\showDOI{\tempurl}


\bibitem[\protect\citeauthoryear{Song, Goh, Fu, Meng, and Heng}{Song
  et~al\mbox{.}}{2011}]%
        {planechi}
\bibfield{author}{\bibinfo{person}{Peng Song}, \bibinfo{person}{Wooi~B. Goh},
  \bibinfo{person}{Chi~W. Fu}, \bibinfo{person}{Qiang Meng}, {and}
  \bibinfo{person}{Pheng~A. Heng}.} \bibinfo{year}{2011}\natexlab{}.
\newblock \showarticletitle{{WYSIWYF: exploring and annotating volume data with
  a tangible handheld device}}. In \bibinfo{booktitle}{\emph{Proc. of
  conference on Human Factors in Computing Systems}}
  \emph{(\bibinfo{series}{CHI '11})}. \bibinfo{publisher}{ACM},
  \bibinfo{pages}{1333--1342}.
\newblock
\showISBNx{978-1-4503-0228-9}
\urldef\tempurl%
\url{https://doi.org/10.1145/1978942.1979140}
\showDOI{\tempurl}


\bibitem[\protect\citeauthoryear{Stoakley, Conway, and Pausch}{Stoakley
  et~al\mbox{.}}{1995}]%
        {wim}
\bibfield{author}{\bibinfo{person}{Richard Stoakley},
  \bibinfo{person}{Matthew~J. Conway}, {and} \bibinfo{person}{Randy Pausch}.}
  \bibinfo{year}{1995}\natexlab{}.
\newblock \showarticletitle{Virtual reality on a WIM: interactive worlds in
  miniature}. In \bibinfo{booktitle}{\emph{Proceedings of the SIGCHI conference
  on Human factors in computing systems}} \emph{(\bibinfo{series}{CHI '95})}.
  \bibinfo{publisher}{ACM Press/Addison-Wesley Publishing Co.},
  \bibinfo{address}{New York, NY, USA}, \bibinfo{pages}{265--272}.
\newblock
\showISBNx{0-201-84705-1}
\urldef\tempurl%
\url{https://doi.org/10.1145/223904.223938}
\showDOI{\tempurl}


\end{thebibliography}
